\documentclass{PoS2}

\usepackage{amsmath}
\usepackage{float}

\title{Massive three loop form factors in the planar limit\thanks{DESY 18--118, DO--TH 18/14} }

\ShortTitle{Massive three loop form factors in the planar limit}

\author{Jakob Ablinger\\
        Research Institute for Symbolic Computation (RISC),\\
        Johannes Kepler University, Altenbergerstra{\ss}e 69, A--4040, Linz, Austria\\
        E-mail: \email{Jakob.Ablinger@risc.jku.at}}

\author{Johannes Bl\"umlein\\
        Deutsches Elektronen--Synchrotron (DESY),\\
        Platanenallee 6, D-15738 Zeuthen, Germany\\
        E-mail: \email{Johannes.Bluemlein@desy.de}}

\author{Peter Marquard\\
        Deutsches Elektronen--Synchrotron (DESY),\\
        Platanenallee 6, D-15738 Zeuthen, Germany\\
        E-mail: \email{Peter.Marquard@desy.de}}

\author{\speaker{Narayan Rana}\\
        Deutsches Elektronen--Synchrotron (DESY),\\
        Platanenallee 6, D-15738 Zeuthen, Germany\\
        E-mail: \email{Narayan.Rana@desy.de}}

\author{Carsten Schneider\\
        Research Institute for Symbolic Computation (RISC),\\
        Johannes Kepler University, Altenbergerstra{\ss}e 69, A--4040, Linz, Austria\\
        E-mail: \email{Carsten.Schneider@risc.jku.at}}

\abstract{
We present the color planar and complete light quark QCD contributions to the three loop heavy quark form factors 
in the case of vector, axial-vector, scalar and pseudo-scalar currents. We evaluate the master integrals applying 
a new method based on differential equations for general bases, which is applicable for any first order factorizing 
systems. The analytic results are expressed in terms of harmonic polylogarithms and real-valued cyclotomic harmonic polylogarithms.
}

\FullConference{Loops and Legs in Quantum Field Theory (LL2018)\\
		29 April 2018 - 04 May 2018\\
		St. Goar, Germany}

\def\asr{\left( \frac{\alpha_s}{4 \pi} \right)} 
\def\b0{\beta_0}

\newcommand{\ep}{\varepsilon}

\begin{document}

\section{Introduction}
\noindent
The top quark, being the heaviest particle of the Standard Model (SM), plays a significant role
in understanding the electro-weak symmetry breaking (EWSB). Besides, its heftiness generates 
a strong potential for hidden beyond the SM (BSM) physics scenarios. Hence, a detailed study 
on top quark observables is always a crucial topic. On the other hand, the abundance of top quark pair 
production at the high energy colliders allows us to obtain accurate measurements.
Especially at the future linear or circular electron-positron colliders, the experimental 
accuracy for this channel will reach ultimate precision. In order to match the experimental
accuracy, precise predictions are required on the theoretical side as well. 
Perturbative quantum chromodynamics (QCD) effects constitute the major contributions
in precision physics and one of the main ingredients of QCD corrections is the form factor.
Form factors are the matrix elements of local composite operators between physical states. 
% In perturbative Quantum Chromodynamics (QCD) these objects play a significant role in determining physical observables. 
In scattering cross-sections, they provide important contributions to the virtual corrections. 
The vector and axial-vector massive form factors are of importance for the forward-backward asymmetry of 
bottom or top quark pair production at electron-positron colliders while, the scalar and pseudo-scalar 
ones may shed light on the decay of a Higgs boson to a pair of heavy quarks.
% and to static quantities like the 
% anomalous magnetic moment of a heavy quark and other processes. 
They are also important to inspect the 
properties of the top quark \cite{Abe:1995hr,D0:1995jca} during the high luminosity phase of the LHC \cite{HLHC} 
and the experimental precision studies at future high energy $e^+ e^-$ colliders \cite{Accomando:1997wt}.

In this note, we present both the color--planar and complete light quark non-singlet three-loop 
contributions to the
massive form factors for vector, axial-vector, scalar and pseudo-scalar currents. Our results except for the vector current,
are presented in \cite{Ablinger:2018yae} and for the vector current,
including the technical details, will be presented elsewhere \cite{FORMF2}. 
In \cite{Bernreuther:2004ih,Bernreuther:2004th,Bernreuther:2005rw,Bernreuther:2005gw},
the two-loop QCD corrections to  the massive vector, axial-vector form factors, the anomaly 
contributions, and the scalar and pseudo-scalar form factors were first presented. In \cite{Gluza:2009yy}, an 
independent computation led to a cross-check of the vector form factor, including the additional 
${\mathcal O}(\ep)$ terms in the dimensional parameter $\ep = (4-D)/2$. The contributions up to 
${\mathcal O}(\ep^2)$ for all the massive two-loop form factors were obtained recently in Ref.~\cite{Ablinger:2017hst}.
The color--planar contributions to the massive three-loop vector form factor have been computed 
in 
\cite{Henn:2016tyf,Henn:2016kjz} and the complete light quark contributions in \cite{Lee:2018nxa}.  
In a parallel and independent computation in \cite{Lee:2018rgs}, the authors also have obtained
both the color--planar and complete light quark non-singlet three-loop massive form 
factors for the aforementioned currents.
In \cite{Grozin:2017aty}, the large $\beta_0$ limit has been considered.

\section{Notation}
\noindent
The notations follow those used in Ref.~\cite{Ablinger:2018yae, Ablinger:2017hst}. 
To summarize, we consider the decay of a virtual massive boson 
of momentum $q$ into a pair of heavy quarks of mass $m$, momenta $q_1$ and $q_2$ and color
$c$ and $d$, through a vertex indicated by $I=V,A,S,P$
for a vector, an axial-vector, a
scalar and a pseudo-scalar boson, respectively.  Here 
$q^2 = (q_1+q_2)^2$ is the center of mass
energy squared and the dimensionless variable $s$ is defined by
%-------------------------------------------------------------------------------------------------------------
\begin{equation}
 s = \frac{q^2}{m^2}\,.
\end{equation}
By studying the Lorentz structure, the following general form of the amplitudes for the vector and axial-vector 
currents can be obtained
\begin{align}
 -i \delta_{cd} ~ \bar{u}_c (q_1) ~ \Big[
 v_Q \Big( \gamma^{\mu} ~F_{V,1} + \frac{i}{2 m} \sigma^{\mu \nu} q_{\nu} ~ F_{V,2}  \Big) 
 +
 a_Q \Big( \gamma^{\mu} \gamma_5~F_{A,1} + \frac{1}{2 m} q^{\mu} \gamma_5 ~ F_{A,2}  \Big) 
 \Big] ~ v_d (q_2) , 
\end{align}
and for the scalar and pseudo-scalar currents
\begin{align}
  - \frac{m}{v} \delta_{cd} ~ \bar{u}_c (q_1) ~ \Big[ s_Q \, F_{S} + i p_Q \gamma_5 \, F_{P} \Big] ~ v_d (q_2) \,.
\end{align}
$\bar{u}_c (q_1)$ and $v_d (q_2)$ are the bi-spinors of the quark and the anti-quark, respectively.
The scalar objects, $F_{I}$, with $I =V, A, S, P$, are the corresponding form factors,
expanded in the strong coupling constant $\alpha_s = g_s^2/(4\pi)$ as follows
%-------------------------------------------------------------------------------------------------------------
\begin{equation}
 F_{I} = \sum_{n=0}^{\infty} \asr^n F_{I}^{(n)} \,.
\end{equation}
$\sigma^{\mu\nu} = \frac{i}{2} [\gamma^{\mu},\gamma^{\nu}]$ and $v_Q, a_Q, s_Q, p_Q$ are the 
vector, axial-vector, scalar and pseudo-scalar coupling constant, respectively.
$v = (\sqrt{2} G_F)^{-1/2}$ is the SM vacuum expectation value of the Higgs field, with $G_F$ being the Fermi constant.
Finally, we multiply appropriate projectors 
as provided in \cite{Ablinger:2017hst}, to obtain the unrenormalized form factors. 
Next, the trace over the color and spinor indices is performed.
For later purposes we denote the number of colors by $N_c$. 
$n_l$ and $n_h$ are the number of light and heavy quarks, respectively.

Since we use dimensional regularization \cite{tHooft:1972tcz}, the important factor for 
axial-vector and pseudo-scalar currents, is a proper definition of $\gamma_5$ in $D$ space-time
dimensions. As both the color-planar and complete light quark contributions belong to the so-called non-singlet case, 
where the axial-vector or pseudo-scalar vertex is connected to open heavy quark lines,
both $\gamma_5$-matrices appear in the same chain of Dirac matrices. Hence we can conveniently use
an anti-commuting $\gamma_5$ in $D$ space-time dimensions, with $\gamma_5^2 = 1$. 
This also implies the well-known Ward identity, 
%-------------------------------------------------------------------------------------------------------------
\begin{equation} \label{eq:cwi}
 q^{\mu} \Gamma_{A,cd}^{\mu, \sf ns} = 2 m \Gamma_{P,cd}^{\sf ns} \,,
\end{equation}
%-------------------------------------------------------------------------------------------------------------
which in terms of the form factors, takes the following form
%-------------------------------------------------------------------------------------------------------------
\begin{equation} \label{eq:cwiFF}
 2 F_{A,1}^{\sf ns} + \frac{s}{2} F_{A,2}^{\sf ns} = 2 F_{P}^{\sf ns} \,.
\end{equation}
%-------------------------------------------------------------------------------------------------------------
Here, the non-singlet contributions are denoted by ${\sf ns}$. For convenience, we introduce 
the Landau variable \cite{Barbieri:1972as}
%-------------------------------------------------------------------------------------------------------------
\begin{equation} \label{eq:varxp}
 x=\frac{\sqrt{q^2-4m^2}-\sqrt{q^2}}{\sqrt{q^2-4m^2}+\sqrt{q^2}}\quad \leftrightarrow 
\quad s = \frac{q^2}{m^2}=-\frac{(1-x)^2}{x}.
\end{equation}
%------------------------------------------------------------------------------------------------------------
% which we use in the following. In particular, we focus on the Euclidean region, $q^2<0$, corresponding to $x \in [0,1[$. 

\section{Computational details}
\begin{figure}[ht]
\begin{center}
    \begin{minipage}[c]{0.13\linewidth}
    \includegraphics[width=1\textwidth]{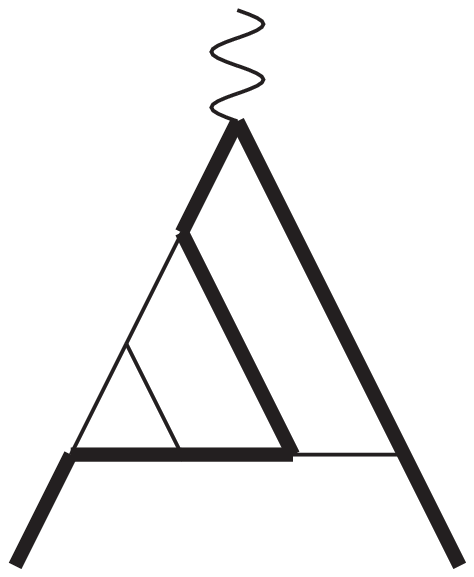}
    \vspace*{-11mm}
    \end{minipage}
        \hspace*{2mm}
    \begin{minipage}[c]{0.13\linewidth}
    \includegraphics[width=1\textwidth]{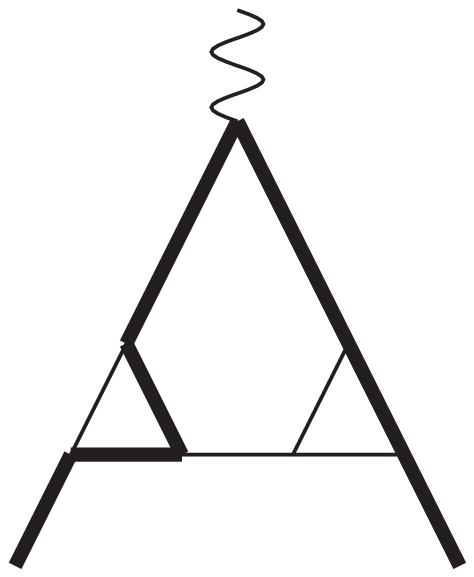}
    \vspace*{-11mm}
    \end{minipage}
        \hspace*{2mm}
    \begin{minipage}[c]{0.13\linewidth}
    \includegraphics[width=1\textwidth]{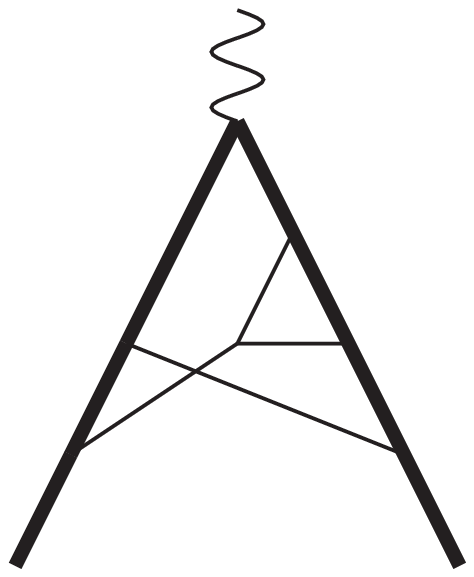}
    \vspace*{-11mm}
    \end{minipage}
\end{center}
\caption{\sf \small The $n_l$ topologies}
\label{fig:nltopo}
\end{figure}

\noindent
We follow the generic procedure to compute the form factors. 
The Feynman diagrams are generated using {\tt QGRAF} \cite{Nogueira:1991ex}.
The {\tt QGRAF} output is then processed 
using {\tt Q2e/Exp} \cite{Harlander:1997zb,Seidensticker:1999bb} and {\tt FORM} \cite{Vermaseren:2000nd, Tentyukov:2007mu}.
The color algebra has been performed using {\tt Color} \cite{vanRitbergen:1998pn}.
By decomposing the dot products among the loop and external momenta, 
the diagrams can be expressed in terms of a linear combination of a large set of scalar integrals.  
% 
%%%%%%%%%%%%%%%%%%%%%%%%%%%%%%%%%%%%%%%%%%%%%%%%%%%%%%%%%%%%%%%%%%%%%%%%%%%%%%%%%%%%%%%%%%%%%%%%%%%%%%%%%%%%%%%%%%%
\begin{figure}[ht]
\begin{center}
\begin{minipage}[c]{0.09\linewidth}
     \includegraphics[width=1\textwidth]{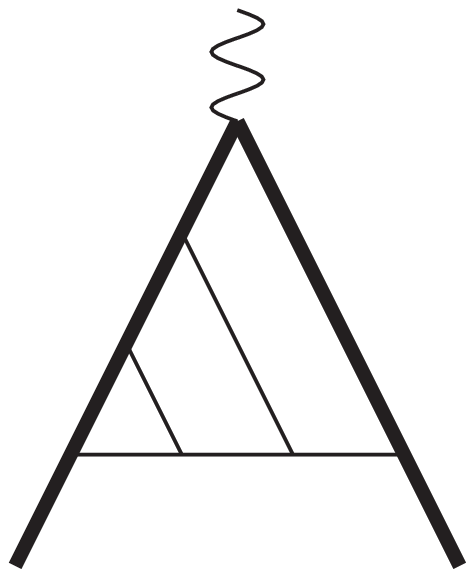}
\vspace*{-11mm}
\begin{center}
% {\footnotesize (a)}
\end{center}
\end{minipage}
\hspace*{2mm}
\begin{minipage}[c]{0.09\linewidth}
     \includegraphics[width=1\textwidth]{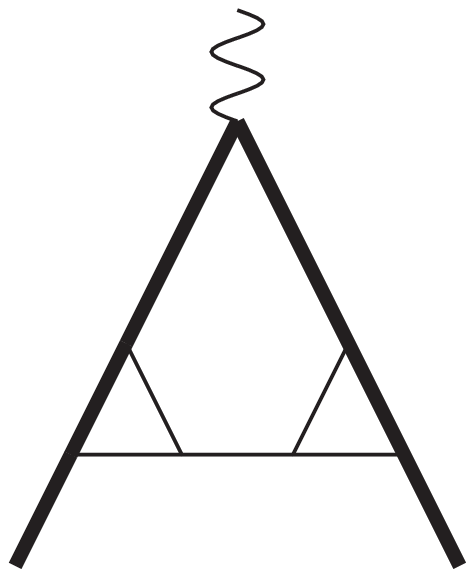}
\vspace*{-11mm}
\begin{center}
% {\footnotesize (b)}
\end{center}
\end{minipage}
\hspace*{2mm}
\begin{minipage}[c]{0.09\linewidth}
     \includegraphics[width=1\textwidth]{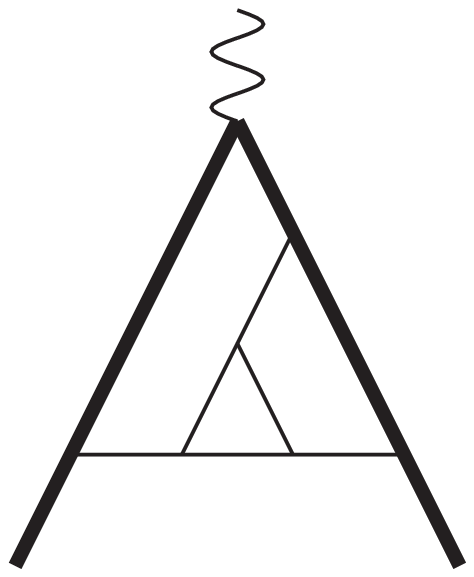}
\vspace*{-11mm}
\begin{center}
% {\footnotesize (c)}
\end{center}
\end{minipage}
\hspace*{2mm}
\begin{minipage}[c]{0.09\linewidth}
     \includegraphics[width=1\textwidth]{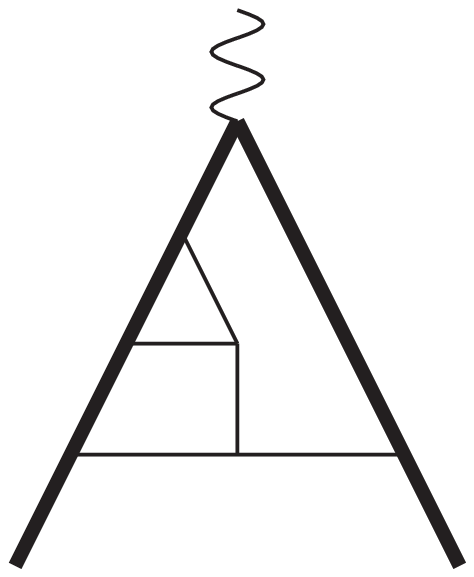}
\vspace*{-11mm}
\begin{center}
% {\footnotesize (a)}
\end{center}
\end{minipage}
\hspace*{2mm}
\begin{minipage}[c]{0.09\linewidth}
     \includegraphics[width=1\textwidth]{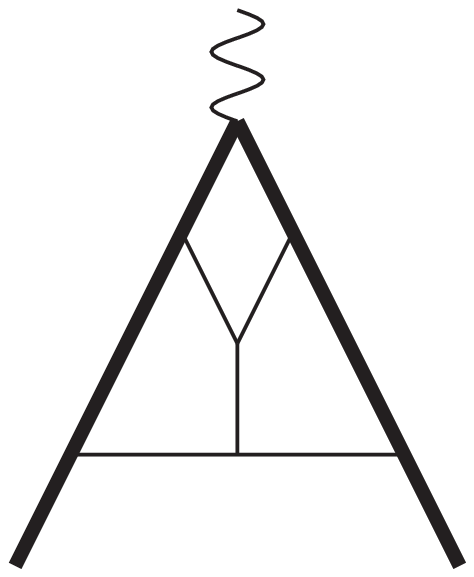}
\vspace*{-11mm}
\begin{center}
% {\footnotesize (b)}
\end{center}
\end{minipage}
\hspace*{2mm}
\begin{minipage}[c]{0.09\linewidth}
     \includegraphics[width=1\textwidth]{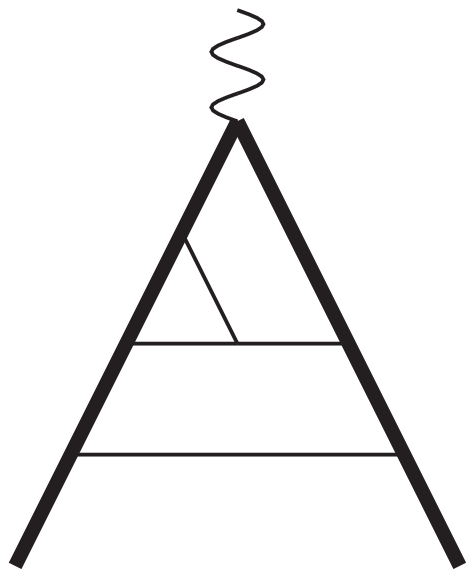}
\vspace*{-11mm}
\begin{center}
% {\footnotesize (b)}
\end{center}
\end{minipage}
\hspace*{2mm}
\begin{minipage}[c]{0.09\linewidth}
     \includegraphics[width=1\textwidth]{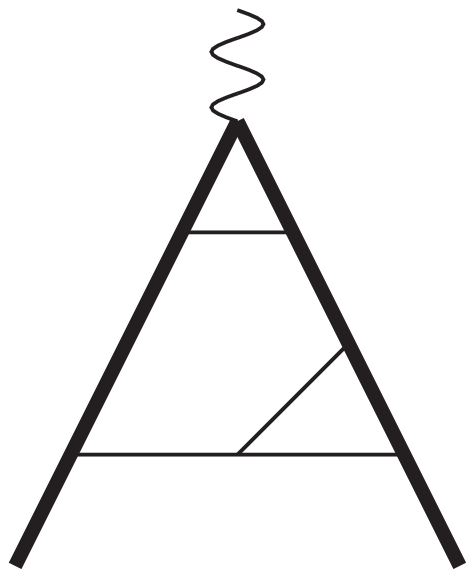}
\vspace*{-11mm}
\begin{center}
% {\footnotesize (b)}
\end{center}
\end{minipage}
\hspace*{2mm}
\begin{minipage}[c]{0.09\linewidth}
     \includegraphics[width=1\textwidth]{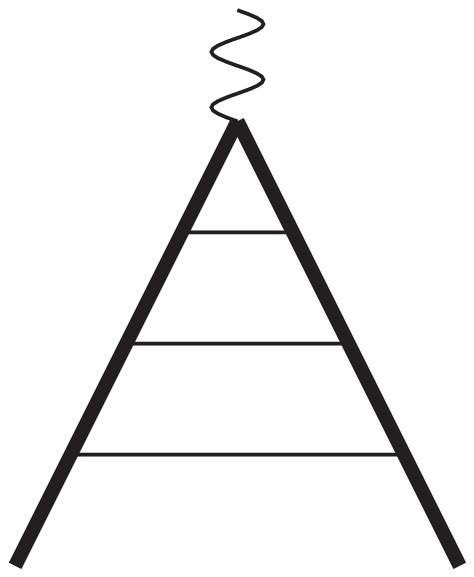}
\vspace*{-11mm}
\begin{center}
% {\footnotesize (b)}
\end{center}
\end{minipage}
\end{center}
\caption{\sf \small The color-planar topologies}
\label{fig:cptopologies}
\end{figure}
%%%%%%%%%%%%%%%%%%%%%%%%%%%%%%%%%%%%%%%%%%%%%%%%%%%%%%%%%%%%%%%%%%%%%%%%%%%%%%%%%%%%%%%%%%%%%%%%%%%%%%%%%%%%%%%%%%%
% 
These integrals are then reduced using integration by parts identities (IBPs)
\cite{Chetyrkin:1981qh,Laporta:2001dd} with the help of the program {\tt Crusher} \cite{CRUSHER} to obtain
109 master integrals (MIs), out of which 96 appear in the color-planar case. 
In the color-planar limit, the families of integrals can be represented by eight topologies, shown in
Figure~\ref{fig:cptopologies}, whereas for the complete light quark contributions, three more topologies, 
cf.~Figure~\ref{fig:nltopo}, are required \footnote{Only sub-topologies with a maximum of eight propagators 
contribute.}.
%%%%%%%%%%%%%%%%%%%%%%%%%%%%%%%%%%%%%%%%%%%%%%%%%%%%%%%%%%%%%%%%%%%%%%%%%%%%%%%%%%%%%%%%%%%%%%%%%%%%%%%%%%%%%%%%%%%
% 

Finally, to compute the MIs, we use the method of differential equations \cite{Kotikov:1990kg,Remiddi:1997ny,Henn:2013pwa,Ablinger:2015tua}. 
For a recent review on the computational methods of loop integrals in quantum field theory, see \cite{Blumlein:2018cms}.
The basic idea is to obtain a set of differential equations of the MIs by performing differentiation
w.r.t $x$ and then to use the IBP relations. 
The first step to solve the corresponding linear system of differential equations
is to find out whether the system is first order factorizable or not. Using the 
package {\tt Oresys} \cite{ORESYS}, based on Z\"urcher's algorithm \cite{Zuercher:94,NewUncouplingMethod}, 
we have found that the present system is indeed first order factorizable in $x$-space. Without any need to 
choose a special basis, we can now simply solve the system in terms of iterated integrals of 
whatsoever alphabet, cf.~Ref.~\cite{FORMF2} for details. The differential equations are solved order by order
in $\varepsilon$ successively, starting at the leading pole terms $\propto 1/\varepsilon^3$. The successive 
solutions in $\varepsilon$ contribute to the inhomogeneities in the next order. We compute 
the master integrals block-by-block, where for an $m \times m$ system, $l$ single inhomogeneous ordinary 
differential equations are obtained, where $1\leq l \leq m$. The orders of these differential equations are
$m_1, \ldots, m_l$ such that $m_1 + \cdots + m_l = m$.
We have solved these differential equations using the variation of constant.
The other $m-l$ solutions result from the former solution immediately. 
The constants of integration are determined using boundary conditions at $x = 1$. 
The calculation is performed by intense use of {\tt HarmonicSums}
\cite{Ablinger:2011te, HSUM,
Ablinger:2014rba, Ablinger:2010kw, Ablinger:2013hcp, Ablinger:2013cf, Ablinger:2014bra}, 
which uses the package {\tt Sigma} 
\cite{Schneider:sigma1,Schneider:sigma2}. Finally, all the MIs
have been checked numerically using {\tt FIESTA} \cite{Smirnov:2008py, Smirnov:2009pb, Smirnov:2015mct}.

The non-homogeneous contributions contain only rational functions of $x$ and hence the results can be 
written in terms of iterative integrals. While integration over a letter is a straightforward algebraic 
manipulation, often $k$-th powers of a letter, $k \in \mathbb{N}$, appear which needs to be transformed
to the letter by partial integration. This method is partially related to the method of 
hyperlogarithms~\cite{Brown:2008um,Ablinger:2014yaa}. We obtain up to weight {\sf w=6} 
real-valued iterated integrals over the alphabet
%------------------------------------------------------------------------------------------------------------------
\begin{eqnarray}
\frac{1}{x},~~
\frac{1}{1-x},~~
\frac{1}{1+x},~~
\frac{1}{1-x+x^2},~~
\frac{x}{1-x+x^2},
\end{eqnarray}
%------------------------------------------------------------------------------------------------------------------
i.e.~the usual harmonic polylogarithms (HPLs) \cite{Remiddi:1999ew} and their cyclotomic extension 
\cite{Ablinger:2011te}, including the respective constants in the limit $x \rightarrow 1$, i.e. the multiple 
zeta values (MZVs) \cite{Blumlein:2009cf} and the cyclotomic constants \cite{Ablinger:2011te,Ablinger:2017tqs,Ablinger:2018xyz}.
% In case of the iterated integrals we apply the linear representation. 
% For a numerical implementation 
The use of shuffle algebra \cite{Blumlein:2003gb}, implemented in {\tt HarmonicSums}, 
reduces the number of functions accordingly, which facilitates numerical evaluation. 
In the MZVs and cyclotomic cases, there are proven reduction relations
to weight {\sf w = 12} \cite{Blumlein:2009cf} and {\sf w = 6} \cite{Ablinger:2017tqs, Ablinger:2018xyz}, 
respectively, which 
have been used. The 188 cyclotomic constants which appear up to {\sf w = 6}, reduce to 23 constants. 
Note that there are more conjectured relations, cf.~\cite{Henn:2015sem}, based 
on PSLQ \cite{PSLQ}. If these conjectured relations are used, only MZVs
remain as constants in all form factors.
% using our real representation for the cyclotomic harmonic polylogarithms.
The analytic result for the different form factors in terms of HPLs and cyclotomic HPLs 
\cite{Remiddi:1999ew,Ablinger:2011te} can be analytically continued  outside $x~\in~[0,1[$ by using the mappings 
$x \rightarrow -x, x \rightarrow (1-x)/(1+x)$ on the expense of extending the cyclotomy class in cases needed.

\section{Ultraviolet renormalization and universal infrared structure}
\noindent
To perform the ultraviolet (UV) renormalization of the form factors, we choose a mixed scheme.
The heavy quark mass and wave function have been renormalized in the on-shell (OS)
renormalization scheme.
We renormalize the strong coupling constant  
in the $\overline{\rm MS}$ scheme, by setting the universal factor $S_\varepsilon = 
\exp(-\varepsilon (\gamma_E - \ln(4\pi))$ for each loop order to one at the end of the calculation.
The required renormalization constants are already well known and denoted by 
$Z_{m, {\rm OS}}$ \cite{Broadhurst:1991fy, Melnikov:2000zc,Marquard:2007uj,
Marquard:2015qpa,Marquard:2016dcn}, 
$Z_{2,{\rm OS}}$ \cite{Broadhurst:1991fy, Melnikov:2000zc,Marquard:2007uj,Marquard:2018rwx} and 
$Z_{a_s}$ \cite{Tarasov:1980au,Larin:1993tp}
for the heavy quark mass, wave function and strong coupling constant, respectively. 
For all the currents, the renormalization of the heavy-quark wave function and the
strong coupling constant are multiplicative, while the renormalization of massive fermion 
lines has been taken care of by properly considering the counter terms.
For the scalar and pseudo-scalar currents, presence of the heavy quark mass in the Yukawa coupling
employs another overall mass renormalization constant, which also has been performed in OS renormalization scheme.

The universal behavior of infrared (IR) singularities of the massive form factors was first
investigated in \cite{Mitov:2006xs} considering the high energy limit. Later in \cite{Becher:2009kw},
a general argument was provided to factorize the IR singularities as a multiplicative renormalization
constant. Its structure is constrained by the renormalization group equation (RGE), as follows,
%-------------------------------------------------------------------------------------------------------------
\begin{equation}
 F_{I} = Z (\mu) F_{I}^{\mathrm{fin}} (\mu)\, ,
\end{equation}
%-------------------------------------------------------------------------------------------------------------
where $F_{I}^{\mathrm{fin}}$ is finite as $\ep \rightarrow 0$. The RGE for $Z(\mu)$ reads
%-------------------------------------------------------------------------------------------------------------
\begin{equation} \label{eq:rgeZ}
 \frac{d}{d \ln \mu} \ln Z(\ep, x, m, \mu)  = - \Gamma (x,m,\mu) \,,
\end{equation}
%-------------------------------------------------------------------------------------------------------------
where $\Gamma$ is the corresponding cusp anomalous dimension, which is by now available up to three-loop 
order \cite{Grozin:2014hna,Grozin:2015kna}. Notice that $Z$ does not carry any information regarding the vertex. 
Both $Z$ and $\Gamma$ can be expanded in a perturbative series in $\alpha_s$ as follows
%-------------------------------------------------------------------------------------------------------------
\begin{equation}
 Z = \sum_{n=0}^{\infty} \asr^n Z^{(n)} \,, \qquad
 \Gamma = \sum_{n=0}^{\infty} \asr^{n+1} \Gamma_{n}
\end{equation}
%-------------------------------------------------------------------------------------------------------------
and one finds the following solution for Eq.~(\ref{eq:rgeZ})
%-------------------------------------------------------------------------------------------------------------
\begin{align} \label{eq:solnZ}
 Z &= 1 + \asr \Bigg[ \frac{\Gamma_0}{2 \ep} \Bigg] 
   + \asr^2 \Bigg[ \frac{1}{\ep^2} \Big( \frac{\Gamma_0^2}{8} - \frac{\beta_0 \Gamma_0}{4} \Big) + \frac{\Gamma_1}{4 \ep} \Bigg] 
   \nonumber\\
  &+ \asr^3 \bigg[ \frac{1}{\ep^3} \left( \frac{\Gamma_0^3}{48} - \frac{\beta_0 \Gamma_0^2}{8} + \frac{\beta_0^2 \Gamma_0}{6} \right)
                 + \frac{1}{\ep^2} \left( \frac{\Gamma_0 \Gamma_1}{8} - \frac{\beta_1 \Gamma_0}{6} \right) 
                 + \frac{1}{\ep} \left( \frac{\Gamma_2}{6} \right) \bigg]
   + {\cal O} (\alpha_s^4) \,.
\end{align}
%-------------------------------------------------------------------------------------------------------------
Eq.~(\ref{eq:solnZ}) correctly predicts the IR singularities for all massive form factors at the
three-loop level.

\section{Results and checks}
\noindent
We finally obtain the color--planar and the complete light quark non--singlet  ($n_l$) contributions for 
the  three-loop  massive 
form factors for vector, axial-vector, scalar and pseudo-scalar currents. 
The expressions, except for the vector current, are attached as supplemental material along with the publication \cite{Ablinger:2018yae}.
The corresponding results for vector current will be available in \cite{FORMF2}.

\begin{figure}[htb]
\centerline{%
\includegraphics[width=0.49\textwidth]{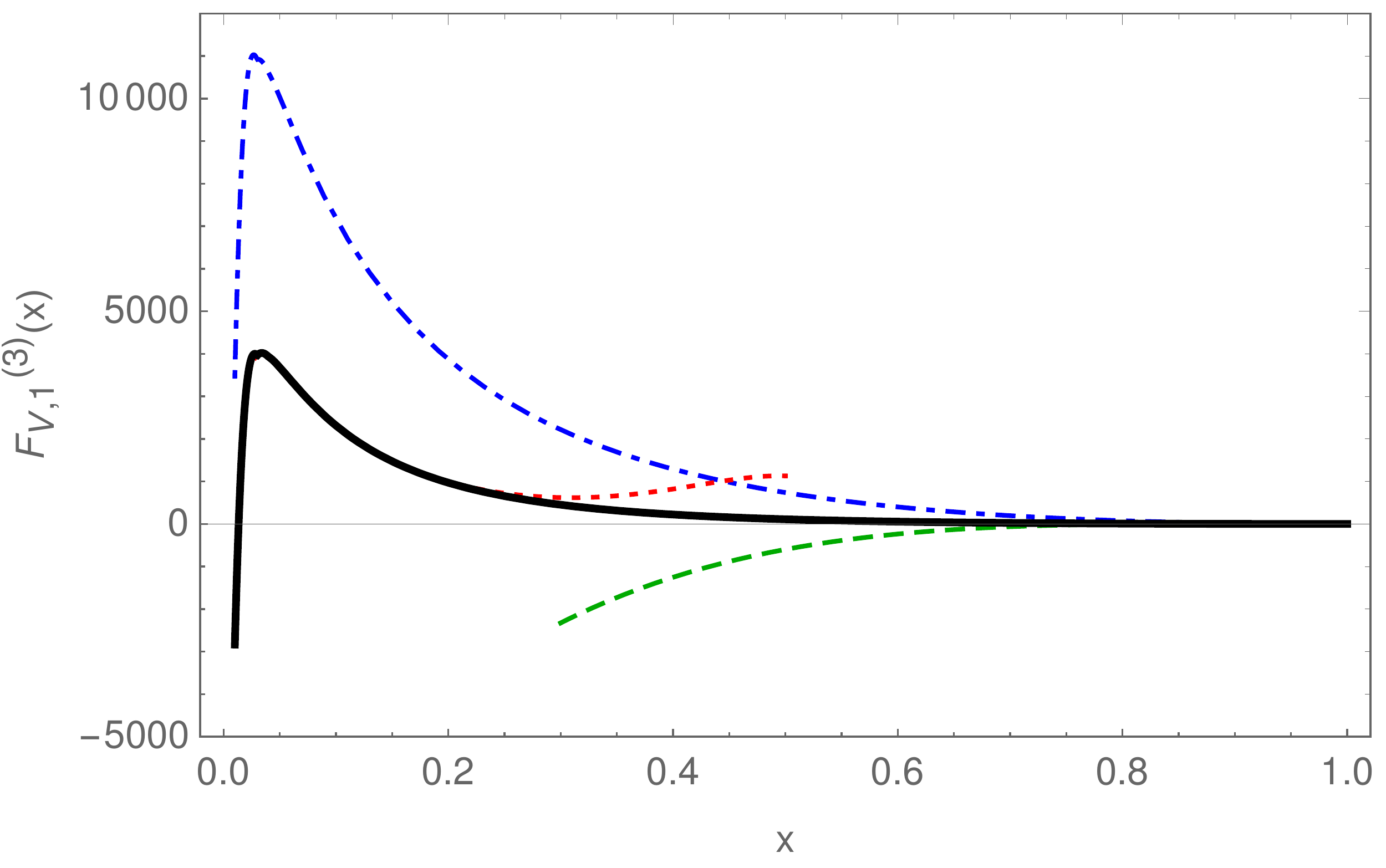}
\includegraphics[width=0.49\textwidth]{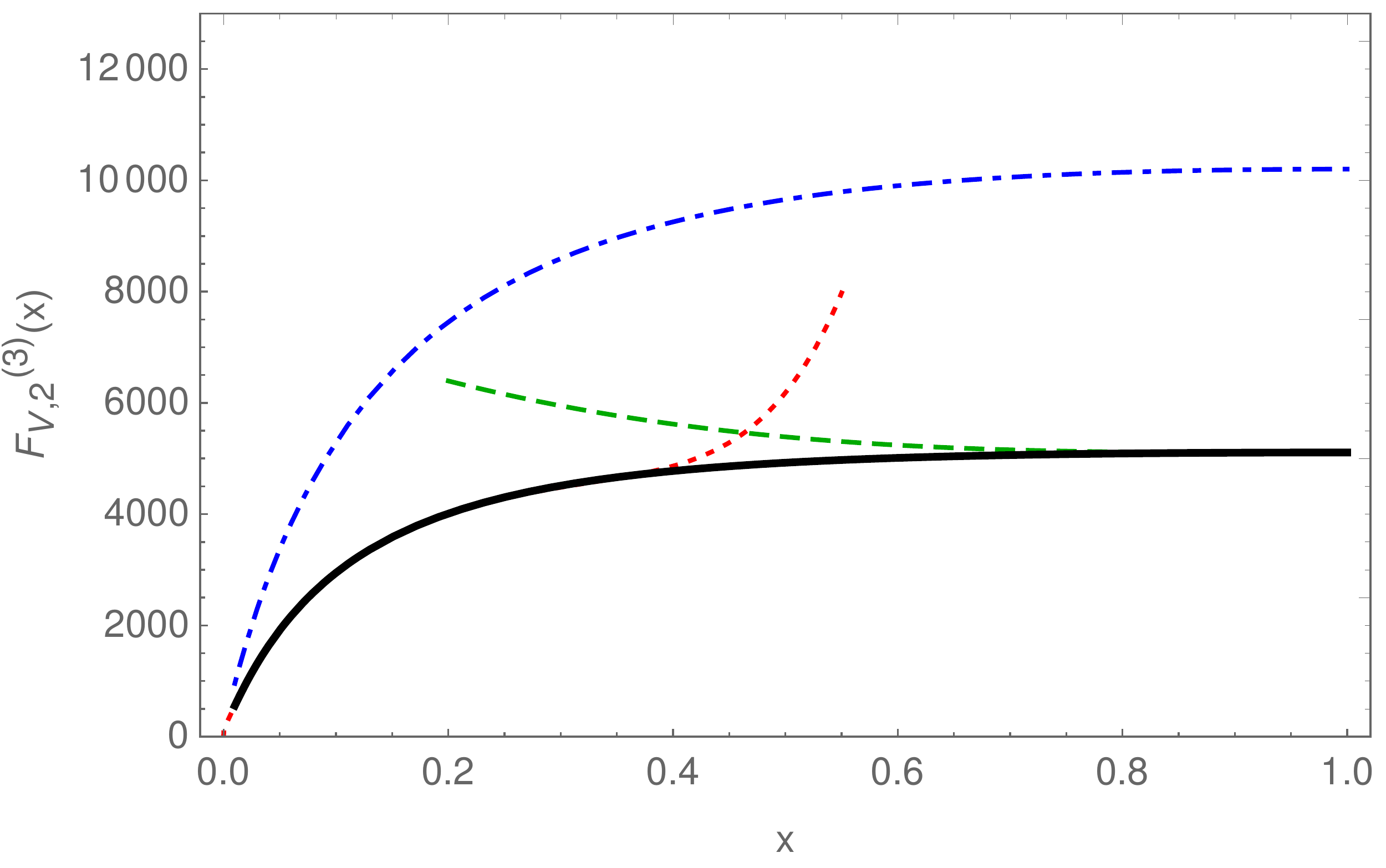}}
\caption{\sf The $O(\varepsilon^0)$ contribution to the vector three-loop form factors 
$F_{V,1}^{(3)}$ (left) and $F_{V,2}^{(3)}$ (right) as a function of $x$. 
Dash-dotted line: leading color 
contribution of the non-singlet form factor; Full line: sum of the complete non-singlet $n_l$-contributions for 
$n_l =5$ and the color-planar non-singlet form factor; Dashed line: large $x$ expansion; Dotted line: small $x$
expansion.}
\label{Fig:VF12}
\end{figure}

\begin{figure}[htb]
\centerline{%
\includegraphics[width=0.49\textwidth]{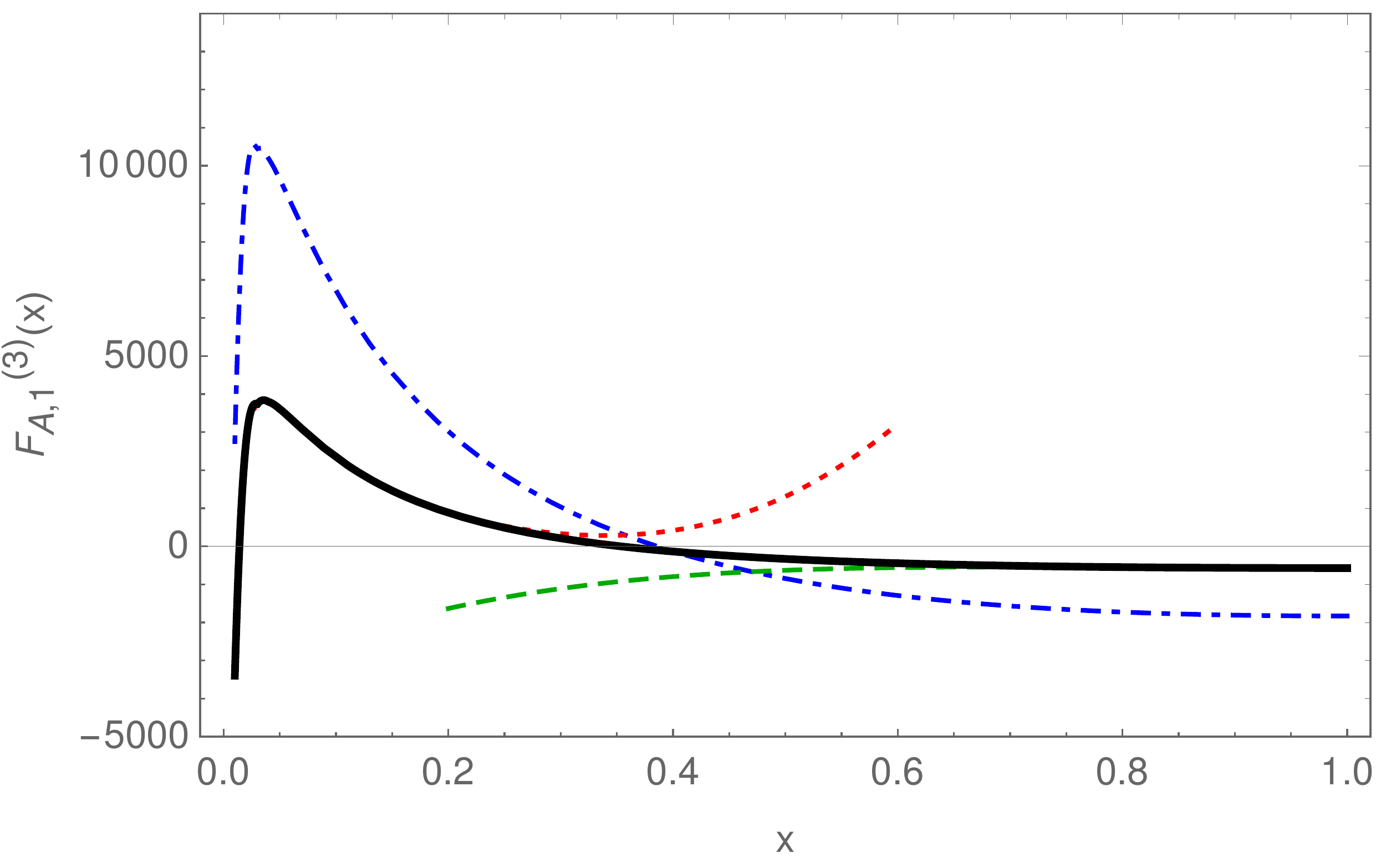}
\includegraphics[width=0.49\textwidth]{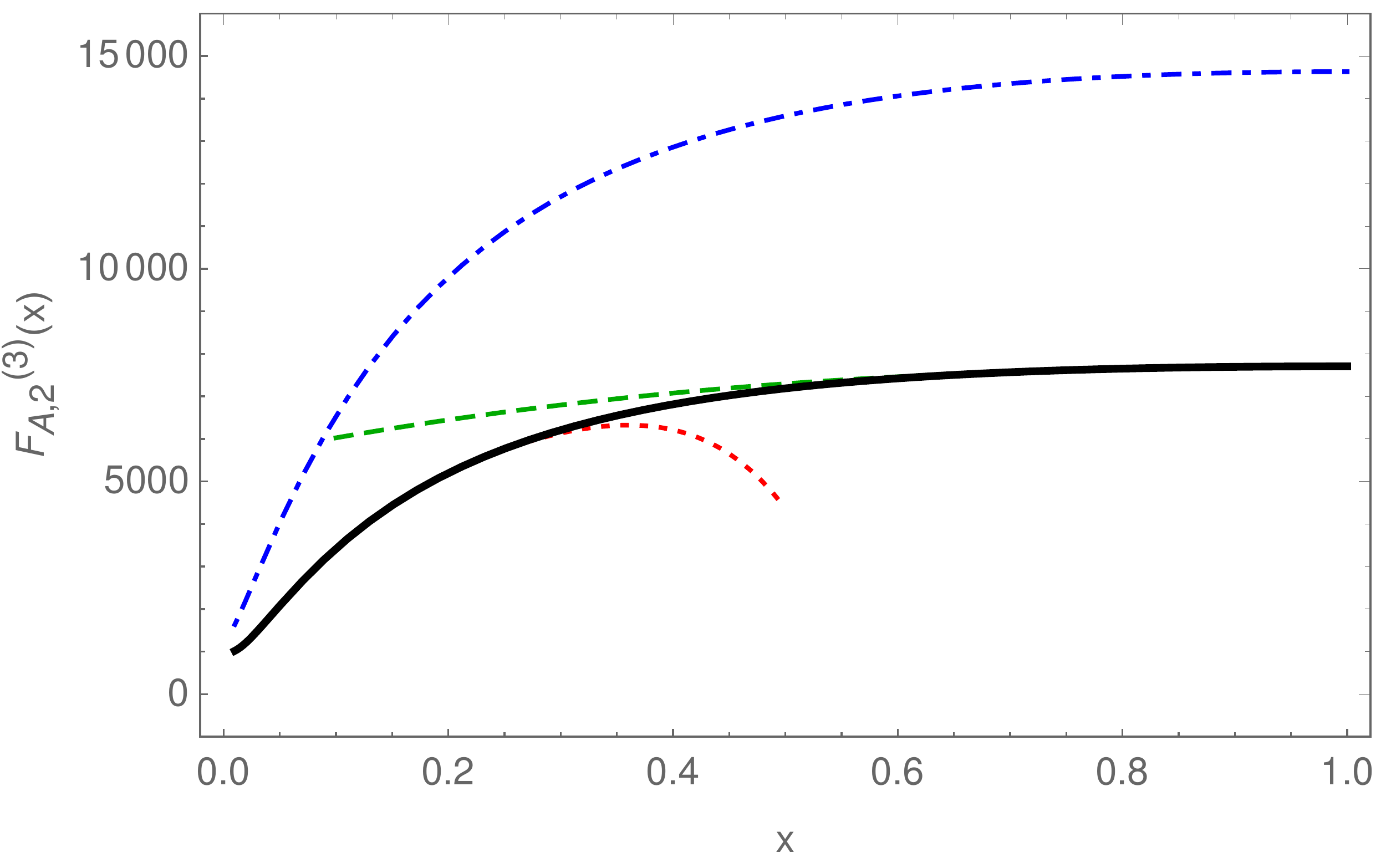}}
\caption{\sf The $O(\varepsilon^0)$ contribution to the axial-vector three-loop form factors 
$F_{A,1}^{(3)}$ (left) and $F_{A,2}^{(3)}$ (right) as a function of $x$. 
Dash-dotted line: leading color 
contribution of the non-singlet form factor; Full line: sum of the complete non-singlet $n_l$-contributions for 
$n_l =5$ and the color-planar non-singlet form factor; Dashed line: large $x$ expansion; Dotted line: small $x$
expansion.}
\label{Fig:AV12}
\end{figure}

\begin{figure}[htb]
\centerline{%
\includegraphics[width=0.49\textwidth]{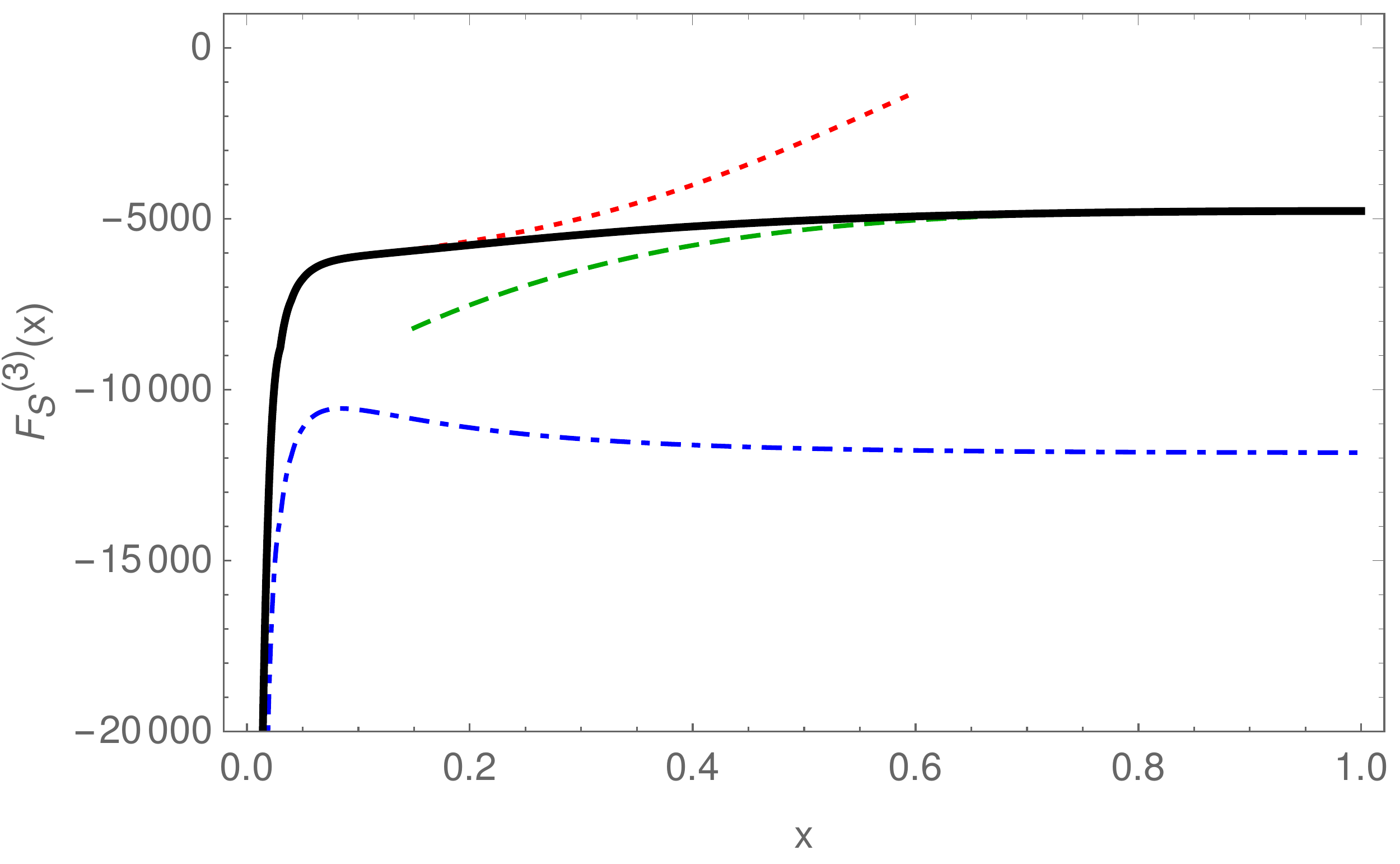}
\includegraphics[width=0.49\textwidth]{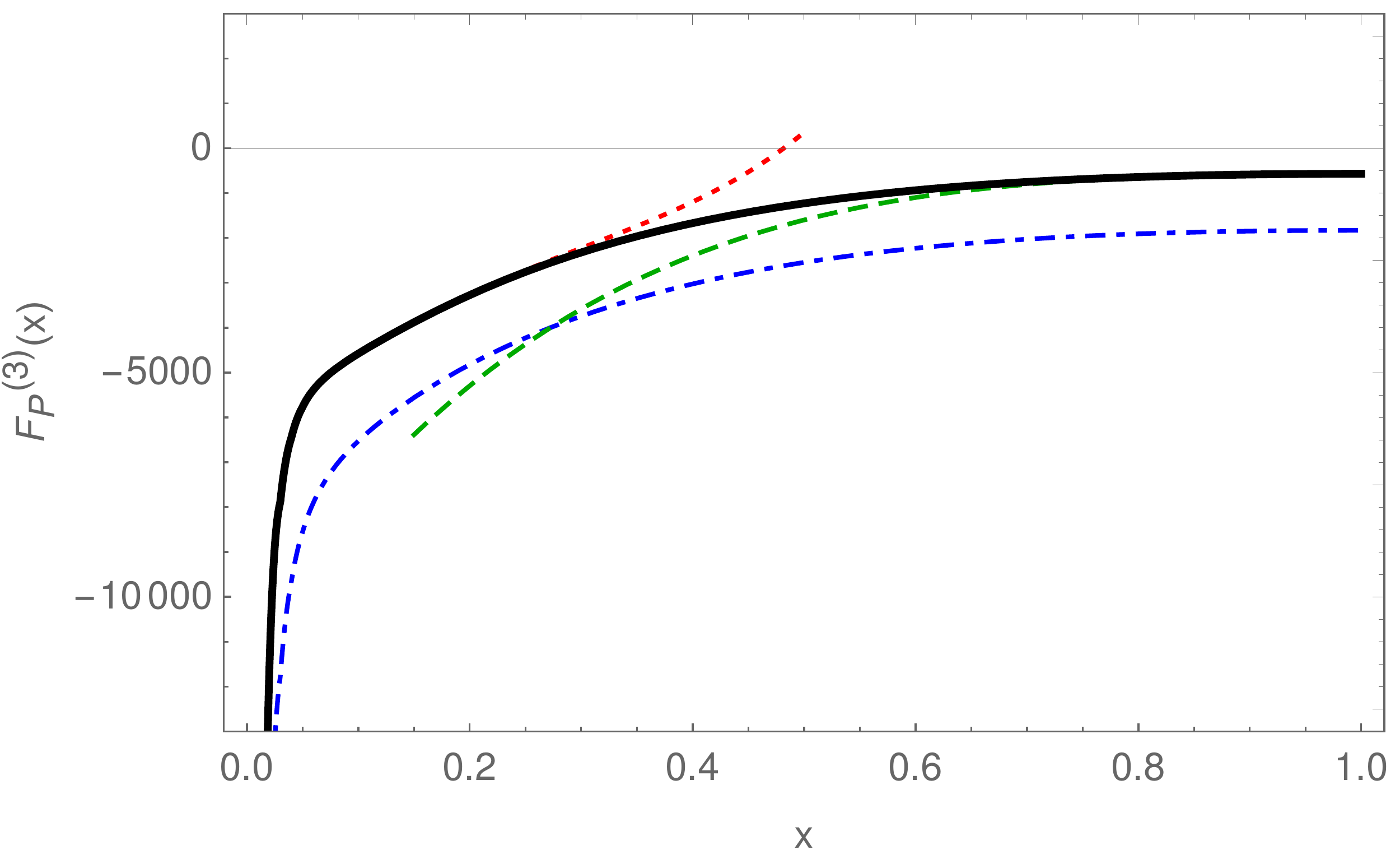}}
\caption{\sf The $O(\varepsilon^0)$ contribution to the scalar and pseudo-scalar three-loop form factors 
$F_S^{(3)}$ (left) and $F_P^{(3)}$ (right) as a function of $x$. 
Dash-dotted line: leading color 
contribution of the non-singlet form factor; Full line: sum of the complete non-singlet $n_l$-contributions for 
$n_l =5$ and the color-planar non-singlet form factor; Dashed line: large $x$ expansion; Dotted line: small $x$
expansion.}
\label{Fig:FSP}
\end{figure}

\noindent
In Figures~\ref{Fig:VF12}--\ref{Fig:FSP} we illustrate the behaviour of the $O(\varepsilon^0)$ 
parts 
of the different form factors
as a function of $x \in [0,1]$. We also show their small- and large-$x$ expansions. The latter 
representations are obtained using {\tt HarmonicSums}. The different limits are characterized as follows :

\noindent 
\emph{Low energy region} ($x \rightarrow 1$): In the space-like case ($q^2 < 0$) we have expanded the
form factors, redefining $x=e^{i\phi}$, $\phi=0$.

\noindent
\emph{High energy region} ($x \rightarrow 0$): 
Here we expand the form factors up to ${\cal O} (x^4)$. The chirality flipping form factors $F_{V,2}$ and 
$F_{A,2}$ vanish and the effect of $\gamma_5$ gets nullified in this limit implying $F_{V,1}=F_{A,1}$ and $F_S=F_P$.
In the small quark mass limit, the form factors satisfy the Sudakov evolution equation.
A detailed study has been performed in \cite{Mitov:2006xs, Ahmed:2017gyt}
to predict part of the vector form factors in this limit from the then available components
up to three and four loop level, respectively.

\noindent
\emph{Threshold region} ($x \rightarrow -1$): We define $\beta = \sqrt{1 - 
\frac{4m^2}{q^2}}$ and expand the form factors around $\beta = 0$.

\noindent
For the numerical evaluation of the HPLs and cyclotomic HPLs in the 
Kummer representation,
we use the {\tt GiNaC}-package
\cite{Vollinga:2004sn,Bauer:2000cp}.

\noindent
We have performed a series of further checks. Through an explicit computation, the Ward identity Eq.~(\ref{eq:cwiFF}) has been checked.
By maintaining the gauge parameter $\xi$ to first order throughout the calculation, a partial 
check on gauge invariance has been achieved. After $\alpha_s$-decoupling, the UV renormalized results satisfy the 
universal IR structure, confirming again the correctness of all pole terms. 
Finally, we have compared our results with those of Ref.~\cite{Lee:2018rgs}, which has been obtained using 
different methods, and agree by adjusting the respective conventions.

\vspace{1ex}
\noindent
{\bf Acknowledgment.}~
This work was supported in part by the Austrian
Science Fund (FWF) grant SFB F50 (F5009-N15). We would like to thank M.~Steinhauser for providing their yet 
unpublished results in electronic form and A.~De Freitas and V.~Ravindran for discussions. The Feynman diagrams 
have been drawn using {\tt Axodraw} 
\cite{Vermaseren:1994je}.

\small
%-------------------------------------------------------------------------------------------------------
\providecommand{\href}[2]{#2}\begingroup\raggedright

%-------------------------------------------------------------------------------------------------------
\endgroup
%-------------------------------------------------------------------------------------------------------
\end{document}